\documentclass[prb,twocolumn,showpacs,preprintnumbers,amsmath,floatfix]{revtex4}
\usepackage{graphicx}
\usepackage{dcolumn}
\usepackage{subfigure}
\usepackage{wrapfig}
\usepackage{cancel}
\usepackage{color}
\begin{document}

\def\k{{\bf k}}
\def\r{{\bf r}}
\def\q{{\bf q}}

\title{ \bf  Lifting of nodes by disorder in extended-$s$ state superconductors: application to ferropnictides}
\author{V. Mishra$^1$, G. Boyd$^1$ S. Graser$^{1,2}$, T. Maier$^3$, P.J. Hirschfeld$^1$, and D.J. Scalapino $^4$}
\affiliation{$^1$University of Florida, Gainesville, FL 32611,
USA\\$^2$ Center for Electronic Correlations and Magnetism,
Institute of Physics, University of Augsburg, D-86135 Augsburg,
Germany\\$^3$ Center for Nanophase Materials Sciences and Computer
Science and Mathematics Division, Oak Ridge National Laboratory,
Oak Ridge, TN 37831-6494\\$^4$ Department of Physics, University
of California, Santa Barbara, CA 93106-9530 USA}
\date{\today}

\begin{abstract}
We show, using a simple model, how ordinary disorder can gap an
extended-$s$ ($A_{1g}$) symmetry superconducting state with nodes.
The concommitant crossover of thermodynamic properties,
particularly the $T$-dependence of the superfluid density, from
pure power law behavior to an activated one is exhibited.  We
discuss applications of this scenario to experiments on the
ferropnictide superconductors.
\end{abstract}

\maketitle

\section{ Introduction}

\label{sec:intro}

When a new class of unconventional superconductors is discovered,
it is traditional to try to determine the symmetry class of the
order parameter, as this may provide clues as to the nature of the
pairing mechanism.  Direct measurements of the superconducting
order parameter are not possible, and indirect phase-sensitive
measurements are sometimes difficult because they typically
involve high-quality surfaces\cite{ref:tsuei}.  On the other hand,
a set of relatively straightforward experimental tests are
available to determine the existence of low-energy quasiparticle
excitations and, in some cases, their distribution in momentum
space. As in earlier experiences with other candidate
unconventional systems like the high-$T_c$ cuprates and heavy
fermion materials, the symmetry class of the newly discovered
ferropnictide superconductors\cite{ref:kamihara} is in dispute at
this writing, due in part to differing results on superfluid
density\cite{ref:Hashimoto,ref:Malone,ref:Martin,ref:Hashimoto2,ref:Gordon,ref:Gordon2,ref:Fletcher},
angle-resolved photoemission
(ARPES)\cite{ref:Zhao,ref:Ding,ref:Kondo,ref:Evtushinsky,ref:Nakayama,ref:Hasan},
nuclear magnetic resonance
(NMR)\cite{ref:RKlingeler,ref:Grafe,ref:Ahilan,ref:Nakai}, Andreev
spectroscopy\cite{ref:Shan,ref:Chien,ref:Daghero,ref:Gonnelli},
 and other probes. In some
cases, results have been taken to indicate the absence of
low-energy excitations, i.e. a fully developed spectral gap. In
others, low-energy power laws have been taken as indication of the
existence of order parameter nodes.  It is not yet clear whether
these differences depend on the stoichiometry or doping of the
materials, or possibly on sample quality.

In parallel, microscopic theoretical calculations of the pairing
interaction in the ferropnictide materials have attempted to
predict the momentum dependence of the order parameter associated
with the leading superconducting instability. Using a 5-orbital
parameterization of the density functional theory (DFT)
bandstructure, Kuroki {\it et al.}~\cite{ref:Kuroki} performed an
RPA calculation of the spin and orbital contributions to the
interaction to construct a linearized gap equation. They found
that the leading pairing instability had $s$-wave ($A_{1g}$)
symmetry, with nodes on the electron-like Fermi surface (``$\beta$
sheets"), and noted that the next leading channel had
$d_{x^2-y^2}$ ($B_{1g}$) symmetry. Wang {\it et
al.}~\cite{ref:Wang}  studied the pairing problem using the
functional renormalization group approach within a 5-orbital
framework, also finding that the leading pairing instability is in
the $s$-wave channel,  and that the next leading channel has
$d_{x^2-y^2}$ symmetry.    For their interaction parameters, they
found however that there were no nodes on the Fermi surface, but
there is a significant variation in the magnitude of the gap.
Other approaches have obtained $A_{1g}$ gaps which change sign
between the hole and electron Fermi surface sheets but remain
approximately isotropic on each
sheet\cite{ref:Mazin_exts,ref:Chubukov_exts}.

 We also recently
presented  calculations of the spin and charge fluctuation pairing
interaction within a 5-orbital RPA framework\cite{ref:graseretal},
using the DFT bandstructure of Cao et al.\cite{ref:Cao} as a
starting point. Our results indicated that the leading pairing
channels were indeed of $s$ ($A_{1g}$) and $d_{x^2-y^2}$ symmetry,
and that one or the other could be the leading eigenvalue,
depending on details of interaction parameters.  We also gave
arguments as to why these channels were so nearly degenerate, and
pointed out some significant differences in the states compared to
those found by Kuroki et al.\cite{ref:Kuroki}.  Finally, we noted
that, within our treatment and for the interaction parameter space
explored, nodes were found in all states, generally on the
$\alpha$ sheets for the $d$-wave case and the $\beta$ sheets for
the $s$-wave case, but that in the latter case the excursion of
the order parameter of sign opposite to the average sign on the
sheet was small, and might be lifted by
disorder\cite{ref:graseretal}.

It is the purpose of this paper to explore the possibility that
the nodes of an extended-$s$ state of the type discussed in Ref.
\onlinecite{ref:graseretal} are lifted by disorder, and consider
the consequences for experimental observables.  In the interest of
simplicity, we initially neglect many complicating aspects of the
problem, in particular the multisheet nature of the Fermi surface,
and focus primarily on the sheet found in each case ($s$ or $d$)
which has nodes.   In this case the problem is similar to one
which was studied earlier in the context of potential extended-$s$
states in a single-band situation for cuprate
superconductors\cite{ref:Borkowski,ref:Fehrenbacher,ref:Muzikar}.
For the pnictides, current interest centers around the isotropic
sign-changing extended-$s$ state proposed by Mazin et
al.\cite{ref:Mazin_exts}, one where a gap is momentum independent
over  two independent Fermi surface sheets, but has a different
sign on each.  In this case, it is known that only interband
scattering is pairbreaking, because it mixes the two signs and
suppresses the overall order
parameter\cite{ref:Muzikar,ref:Mazindisorder,ref:Kontanidisorder,ref:YBang}.
This effect has been claimed to account for low-energy excitations
observed in NMR\cite{ref:MazinNMR,ref:ChubukovNMR} and superfluid
density\cite{ref:Chubukov_rhos} experiments. In general,
impurities with screened Coulomb potential will have a larger
intraband component, which is however essentially irrelevant for
the isotropic case in the sense of Anderson's theorem.  However,
intraband scattering by disorder will average any anisotropy of
the order parameter present in the conventional $s$-wave
case\cite{ref:Kadanoff}. In the case of the pnictides, this
intraband component will do the same, with the effect of lifting
the weak nodes found in our microscopic calculations. We
demonstrate this effect, and its consequences, within a simple
model where we take weak pointlike scatterers, treated in the Born
approximation, as a model for the intraband part of the true
disorder scattering in the pnictides. We believe that such
scattering  arises primarily from out-of plane dopants like K in
the hole-doped and F in the electron-doped materials; since these
dopants sit away from the FeAs plane, they will certainly produce
a significant small-$\bf q$ component of the scattering potential,
which will mix states on the same sheet. Our results suggest that
an extended-$s$ wave state with nodes, lifted by disorder in the
case of the doped superconducting ferropnictides, may explain the
apparent discrepancies among various measurements in the
superconducting states of these materials.

\begin{figure}[]
\includegraphics[width=.85\columnwidth]{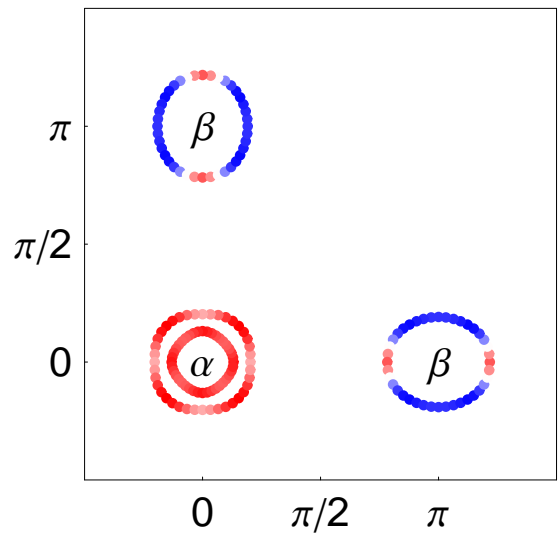}
\caption{(Color online)  Fermi surface and extended-$s$ gap
eigenfunction from Ref. \onlinecite{ref:graseretal}, Fig. 15(d).
Red/blue color indicates sign of order parameter. } \label{fig1}
\end{figure}

\section{Model}
\label{sec:model} We begin by assuming a superconductor with a
separable pair interaction $V(\k,\k')=V_1
\Phi_1(\phi)\Phi_1(\phi')$, where $\phi$ is an angle
parameterizing the electronic momentum $\k$ on a single circular
Fermi surface. To model a situation corresponding to an
extended-$s$ state on the $\beta$ sheet of the Fe-pnictide
materials, as found in Ref. \onlinecite{ref:graseretal}, we choose
\begin{equation}
\Phi_1(\phi)= 1 + r \cos 2\phi, \label{eq:potential}
\end{equation}
with $r\equiv V'/V_1\gtrsim 1$, ensuring that the pure order
parameter $\Delta_\k =\Delta_0\Phi_1(\phi)$ will have nodes near
$\phi=\pi/4, 3\pi/4, ...$.  Note that this function is also
suitable for modelling the $d$-wave state on the $\alpha$ Fermi
surface sheets, in the limit $r\gg 1$.  Because the effect of
disorder will renormalize the constant and $\cos 2\phi$ parts of
the order parameter differently, we write it in the general case
as $\Delta_\k=\Delta_0 + \Delta'\cos 2\phi$.

 The full matrix Green's function in the
presence of scattering in the superconducting state is
\begin{equation}
G({\bf k}, \omega) = {\tilde\omega\tau_0
+\tilde\epsilon_\k\tau_3+\tilde\Delta_{\bf k}\tau_1\over
\tilde\omega^2-\tilde\epsilon_\k^2-\tilde\Delta_\k^2},
\label{Greensfctn}
\end{equation}
where $\tilde\omega\equiv \omega-\Sigma_0$,
$\tilde\epsilon_\k\equiv \epsilon_\k+\Sigma_3$, $\tilde \Delta_\k
\equiv \Delta_\k+\Sigma_1$, and the $\Sigma_\alpha$ are the
components of the self-energy proportional to the Pauli matrices
$\tau_\alpha$ in particle-hole (Nambu) space.  If we  assume  weak
scattering, we may approximate the self-energy in the Born
approximation  as
\begin{equation}
\underline{\Sigma} = n_I \sum_{k^\prime} |U(\k\k')|^2\tau_3
\underline{G}^0(\k',\omega)\tau_3, \label{BornSC}
\end{equation}
where $n_I$ is the concentration of impurities.  The self-energy
has Nambu components
\begin{eqnarray}
\Sigma_0(\k,\omega) & = & {n_I}\,  \sum_{\k'}|U(\k, \k^\prime)|^2
\ \frac{\tilde\omega }{\tilde\omega^2-\tilde\epsilon^2_{\k^\prime}
-
\tilde\Delta^2_{\k^\prime} }\, ,\label{Sig0}\\
\Sigma_3(\k,\omega) & = & {n_I}\,  \sum_{\k'} |U(\k ,\k^\prime)|^2
\
\frac{\tilde\epsilon_{\k^\prime}}{\tilde\omega^2-\tilde\epsilon^2_{\k^\prime}
-
\tilde\Delta^2_{\k^\prime} }\, , \label{Sig3}\\
\noalign{\hbox{and}} \Sigma_1(\k,\omega) & = &  -{n_I}\,
\sum_{\k'} |U(\k, \k^\prime)|^2 \
\frac{\tilde\Delta_{\k^\prime}}{\tilde\omega^2-\tilde\epsilon^2_{\k^\prime}
- \tilde\Delta^2_{\k^\prime} }\, . \label{Sig1}
\end{eqnarray}

 As discussed above, we assume
pointlike impurity scattering $U(\k,\k')=U_0$, and treat the
disorder in Born approximation.  We further assume particle-hole
symmetry such that $\Sigma_3=0$,  leading to  Nambu self-energy
components after integration perpendicular to the Fermi surface

\begin{eqnarray}
 \Sigma_0(\phi,\omega)= \Gamma \left\langle\frac{\tilde{\omega}}{\sqrt{\tilde{\omega}^2-{\tilde\Delta_{\k'}}^2}}\right\rangle_{\phi'}\\
 \Sigma_1(\phi,\omega) = -\Gamma \left\langle\frac{{\tilde\Delta_{\k'}}}{\sqrt{\tilde{\omega}^2-{\tilde\Delta_{\k'}}^2}}\right\rangle_{\phi'}
 ,\label{eq:dirty}
 \end{eqnarray} where $\langle\rangle_\phi$ indicates averaging over the circular Fermi surface, and   $\Gamma=\pi
 n_I N_0
 U_0^2$ is the normal state scattering rate,  with $N_0$ the
 density of states at the Fermi level.

\section{Results}
\label{sec:results}
 The BCS gap equation \begin{equation}
 \Delta_\k = {1\over 2}{\rm Tr}\,T\sum_{\omega_n}\sum_{\k'} V_{\k,\k'} \tau_1 G(\k',\omega_n) \end{equation}
 then reduces to

 \begin{eqnarray}
\Delta_0 &=& N_0 \left[ V_1 \Delta_0 I_1 +(V^\prime  \Delta_0 +
V_1 \Delta^\prime) I_2+
V^\prime \Delta^\prime I_3\right]\nonumber\\
\Delta' &=& N_0 \left[ V^\prime \Delta_0 I_1 +(V^\prime
\Delta^\prime + rV^{\prime } \Delta_0) I_2 + rV^{\prime }
\Delta^\prime I_3\right]\nonumber \\\label{eq:gap_eqn}
\end{eqnarray}
with
\begin{eqnarray}
I_m&=& \pi T \sum_{\omega_n } \left\langle\frac{(\cos 2\phi)^{m-1}}{\sqrt{\tilde\omega_n^2+\tilde\Delta_{\k'}^2}}
\right\rangle_{\phi'},
% \\
%I_2&=& \pi T \sum_{\omega_n } \left\langle\frac{\cos{2 \phi^{\prime}}}{\sqrt{\tilde\omega_n^2+\tilde\Delta_{\k'}^2}} \right\rangle_{\phi'}\\
%I_3&=& \pi T \sum_{\omega_n } \left\langle\frac{(\cos{2 \phi^{\prime}})^2}{\sqrt{\tilde\omega_n^2+\tilde\Delta_{\k'}^2}}\right\rangle_{\phi'} \\
\label{eq:Is}
\end{eqnarray}
and where $\omega_n=(2n+1)\pi T$ is a fermionic Matsubara
frequency.

\begin{figure}[]
\includegraphics[width=.85\columnwidth]{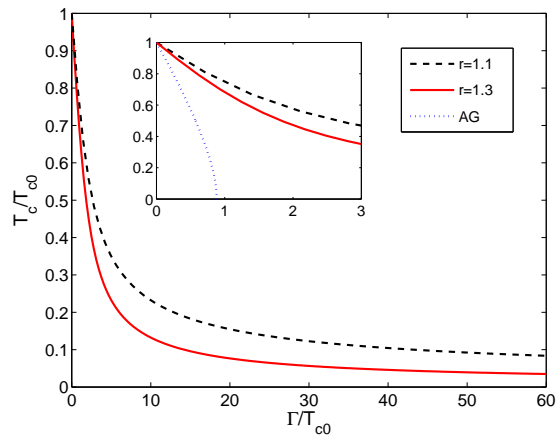}
\caption{(Color online)  Suppression of the critical temperature
$T_c/T_{c0}$ vs.
 normal state scattering rate
$\Gamma/T_{c0}$ for  anisotropy parameter $r=1.1 (dashed),1.3
(solid)$. The inset shows the two curves at smaller values of the
scattering rate. The $T_c$ suppression curve expected from
Abrikosov-Gor'kov theory\cite{ref:AG} (dotted) is included for
comparison.} \label{fig2}
\end{figure}

\subsection{$T_c$ suppression}

Near $T_c$ $ I_1 = {\cal L}$ , $I_2 = 0$ and $I_3 =  {\cal L}/2$ ,
where $ {\cal L} = \log\left[\frac{2 e^{\gamma} \omega_c}{\pi
T_c}\right]$, leading to a critical temperature of

 \begin{equation}
 T_c=1.13 \omega_c \exp \left[-\frac{V_1}{N_0 (V_1^2+V^{\prime 2}/2)} \right]
 \label{eq:Tc}
 \end{equation}
in the clean limit.  In the dirty case, $T_c$ is suppressed by
ordinary disorder until the gap anisotropy is completely washed
out, at values of the normal state scattering rates  $\Gamma$ many
times larger than $T_c$, as also found by Markowitz and
Kadanoff\cite{ref:Kadanoff}.  In Fig. \ref{fig2}, we plot this
behavior by solving (\ref{eq:gap_eqn}) with  $\Phi_1$ given by Eq.
(\ref{eq:potential}).    The marginal case $r=1$ describes the
situation where the nodes just touch the Fermi surface without
disorder, which has been discussed earlier\cite{ref:Borkowski}.
Note that while the scale of the plot implies that the critical
temperature is rapidly suppressed, the scattering rate scale over
which this occurs is actually much greater than one would expect
in, e.g. a $d$-wave superconductor, where a critical concentration
$\Gamma_c\sim T_{c0}$ suffices to suppress superconductivity
completely.

\begin{figure}[]
\includegraphics[width=.85\columnwidth]{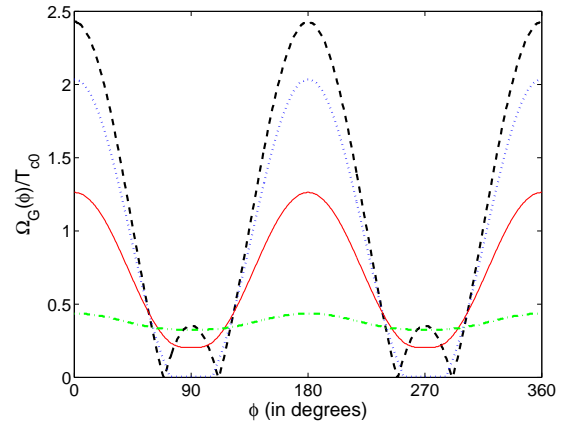}
\caption{(Color online)  Normalized spectral gap
$\Omega_G(\phi)/T_{c0}$ vs. angle $\phi$ on the Fermi surface for
an extended $s$-wave state with $r=1.3$ and $\Gamma/T_{c0}=0$
(dashed), 0.3 (dotted), 1.0 (solid), and 3.1 (dashed-dotted). }
\label{fig3}
\end{figure}

\subsection{Density of states and spectral gap}

In the symmetry broken state, the order parameter $\Delta_\k$ is
renormalized by the off-diagonal self-energy
$\Sigma_1(\k,\omega)$, but the true spectral gap is determined by
both Nambu components $\Sigma_1$ and $\Sigma_0$.   One may define
an angle-dependent spectral gap $\Omega_G(\phi)$ by examining the
1-particle spectral function $A(\k, \omega)\equiv -(1/\pi) {\rm
Im}\, G_{11}(\k,\omega)$, and plotting either the peak or the
energy between the peak and the Fermi level where $A$ falls to
one-half its peak value.   We adopt the latter definition here
since it is similar to one traditionally used in the ARPES
community, but other measures (e.g. plotting the peak in
$A(\k,\omega)$ or simply $\tilde \Delta(\omega=0,\phi)$) produce
very similar results.  In Fig. \ref{fig3}, we show the $s$-wave
spectral gap plotted over the Fermi surface for various values of
the scattering rate $\Gamma$.  It is clear that the effect of
scattering is as described above: the order parameter is averaged
over the Fermi surface, which eventually has the effect of lifting
the nodes, and leading to a true gap in the system, which becomes
isotropic at sufficiently large scattering rates.
\begin{figure}[]
\includegraphics[width=.85\columnwidth]{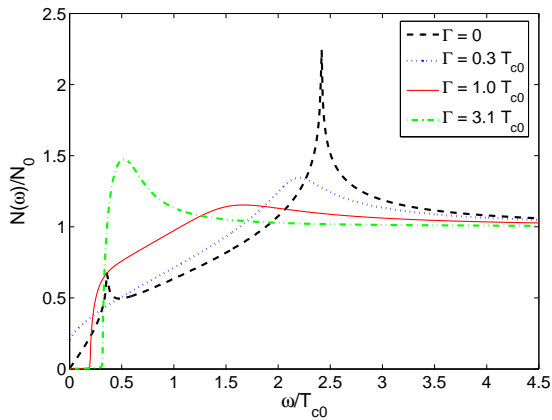}
\caption{(Color online)  Normalized density of states
$N(\omega)/N_0$ vs. energy $\omega$ for the same parameters and
line types as Fig. \ref{fig3}.
  } \label{fig4}
\end{figure}

This can be observed as well in the total density of states, which
we exhibit in Fig. \ref{fig4}.  In the clean situation, there are
two coherence-type peaks, corresponding to the large and small
antinodal order parameter scales observed in Fig. \ref{fig3}. The
addition of disorder smears these peaks initially and suppresses
the maximum gap feature.   This is consistent with the expected
behavior of a dirty nodal superconductor.  This behavior continues
until the point when the nodes are actually lifted, and the system
acquires a true gap with  sharp coherence peaks. When the gap has
become completely isotropic due to disorder averaging, the DOS is
identical to the usual BCS DOS, as required by Anderson's theorem.

\subsection{Superfluid density}
 The temperature
dependence of the superfluid density $\rho_s(T)$ reflects the
distribution of quasiparticle states which contribute to the
normal fluid fraction. Early data on powdered  samples of
R-FeAsO$_x$F$_{1-x}$ (R =Pr\cite{ref:Hashimoto},
Sm\cite{ref:Malone},Nd\cite{ref:Martin}) near optimal doping
showed exponential $T$ dependence, as did measurements on
Ba$_x$K$_{1-x}$Fe$_2$As$_2$ (Ba-122)\cite{ref:Hashimoto2}.   More
recent experiments on the latter system doped with Co found a
power law temperature dependence close to $T^2$ which evolved
towards exponential with increasing Co
concentration\cite{ref:Gordon2}. A $T^2$ dependence is
characteristic of a dirty system with linear
nodes\cite{ref:Gross}, but the authors of Ref.
\onlinecite{ref:Gordon2} concluded the $T^2$ was more likely due
to nonlocal electrodynamics\cite{ref:Koztin}, or pairbreaking due
to inhomogeneity or inelastic scattering.  Very recently, however,
a {\it linear} $T$ dependence was measured in the original
ferropnictide superconductor LaFePO,\cite{ref:KamiharaP} with
$T_c=6K$.  Because this material is stoichiometric, crystals have
very long mean free paths   of more than a thousand Angstr\"om and
are capable of supporting dHVA oscillations\cite{ref:Carrington1}.
It seems very unlikely that anything but order parameter nodes can
lead to such a power law.  The pure $s_{+/-}$ state we consider
here with weak nodes will immediately yield
$\rho_s(0)-\rho_s(T)\sim T$, so we take the parameters of the
previous section and calculate $\rho_s$ directly.

 Within the current BCS-type model, the $xx$ component of the superfluid density tensor $\rho_s$ may be
written

\begin{equation}
\rho_{s,xx}/\rho = \left\langle \cos ^2 \phi \int d\omega \,\tanh
{\omega\over 2 T} \, {\rm Re\,} {\tilde \Delta_\k^2\over (\tilde
\omega^2-\tilde \Delta_\k^2 )^{3/2}}\right\rangle_\phi,
\label{eq:rhos}
\end{equation}
where $\rho$ is the full electron density.
\begin{figure}[]
\includegraphics[width=.85\columnwidth]{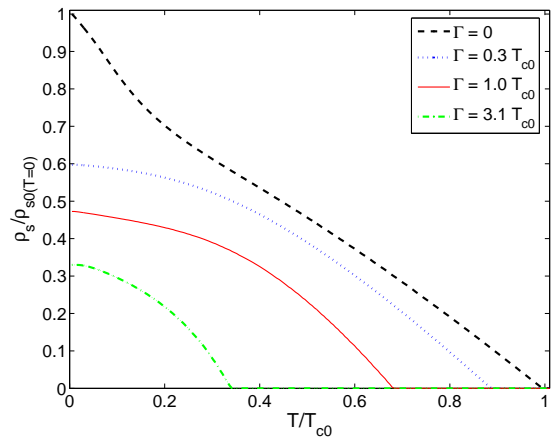}
\caption{(Color online)   Superfluid density $\rho_s/m^*$ vs.
$T/T_{c0}$ for $r=1.3$ and same disorder parameters and line types
as in Figs. \ref{fig3}-\ref{fig4}.  } \label{fig5}
\end{figure}
Note this standard expression for the superfluid density of a
dirty superconductor\cite{ref:Skalski} is complete within mean
field theory since vertex corrections to the current response
vanish for singlet superconductors and pointlike
scatterers\cite{ref:HWE88}. In Fig. \ref{fig5}, we show the
superfluid density ${\overline\rho}_s\equiv
(\rho_{s,xx}+\rho_{s,yy})/2$ calculated from Eq. (\ref{eq:rhos})
vs. temperature for various scattering rates\cite{ref:footnote1}.
In the pure system, the superfluid density is linear at the lowest
temperatures, reflecting the existence of line nodes. At a
temperature corresponding to the smaller antinodal
gap\cite{ref:footnote2}, there is a decrease in the rate at which
thermally excited quasiparticles depopulate the condensate, and
therefore a change in slope as visible in the figure.  We note
that this kinklike behavior may disappear when excitations from
other Fermi surface sheets which contribute at higher temperatures
are included, but also that this type of upward curvature in
$\rho_s$ vs. $T$ was observed by Fletcher et
al.\cite{ref:Fletcher} in the cleanest LaFePO samples.

\begin{figure}[]
\includegraphics[width=.85\columnwidth]{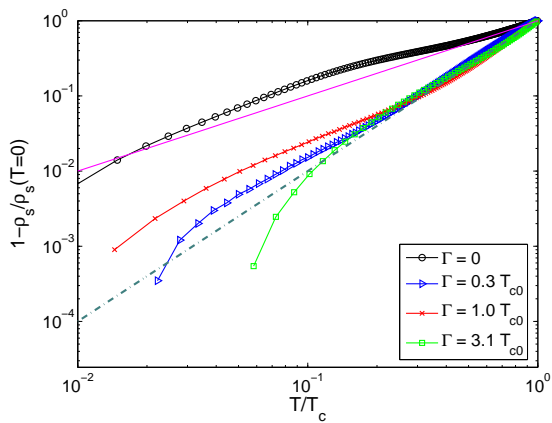}
\caption{(Color online)  $\log_{10} (1-\rho_s(T)/\rho_s(0))$ vs.
$\log_{10} T/T_c$ for $r=1.3$ and various scattering rates
(Circles: $\Gamma=0$, triangles: $\Gamma/T_{c0}=0.3$, crosses:
$\Gamma/T_{c0}=1.$, and squares: $\Gamma/T_{c0}=3.1$ .  Solid
line: $T$; dashed-dotted line: $T^2$.
 }
\label{fig6}
\end{figure}

 If disorder is now added without
lifting the nodes, the generic behavior will be quadratic in
temperature\cite{ref:Gross}, i.e. $\rho_s \simeq \rho_s(0)(1
-aT^2/\Gamma^2)$, where $a$ is a constant of order unity.  When
the nodes are completely lifted, an exponential $T$ dependence,
$\rho_s \simeq \rho_s(0)(1 -a \exp(-\Delta_{min}/T))$ must
dominate at the lowest temperatures. These power laws are
displayed more precisely in the log-log plot of Fig. \ref{fig6}.
The pure system follows a linear-$T$ law ($\rho_s(0)-\rho_s(T)\sim
T$) down to the lowest temperatures, as expected because of the
linear nodes, whereas the dirty systems follow a $T^2$ law over an
intermediate $T$ range, also expected because the states are
uniformly  broadened in this range by Born scattering with
$1/\tau$ roughly linear in energy.   At lower temperatures,
however, the dirty cases no longer exhibit power laws in
temperature, with the curves showing fully developed gaps in Fig.
\ref{fig4} following activated behavior.  The $\Gamma/T_{c0}=0.3$
case is marginal in the sense that the effective spectral gap has
nodes which just touch the Fermi surface, and the figure shows
that when the system is close to this condition the low-$T$
behavior is not a simple power law or exponential.

\section{Effect of additional Fermi surface sheets}
\label{sec:additional}

 While the desired effect of node lifting by
disorder for an extended-$s$ wave state with weak nodes has now
been exhibited, the amount of disorder required to lift the nodes
has also been shown to suppress $T_c$ substantially.  As strong
sample-to-sample variations of $T_c$ have not been observed in the
ferropnictides, this seems inconsistent with the current overall
body of experimental data.  There are several possibilities to
explain this discrepancy.  The first is that the nodes, e.g. in
the LaFePO system, may be accidentally extremely weak.  This seems
unlikely, since the linear-$T$ behavior extends over a significant
fraction of $T_c$ in this material.  Were the nodes to be marginal
or extremely weak, the phase space from near-quadratic nodes would
actually lead to a $T^{1/2}$ temperature variation of the
penetration depth in the pure system, which is not observed. We
have also examined the above model with smaller values of
$r\gtrsim 1$, and find that substantial scattering rates are still
required to create a significant spectral gap.

It is possible that our restriction to Born (weak) scattering is
inadequate.  Unitary scatterers  modify primarily the states near
the nodes, creating a gap without affecting the states at higher
energies; smaller normal state scattering rates can therefore
produce a large node lifting effect without suppressing $T_c$
significantly\cite{ref:felds}.  It may be interesting to explore
this effect further, but a priori there is no obvious reason for
the out-of-plane impurities which appear to dominate the doped
pnictide materials to produce such strong scattering potentials.

A final possibility, which we do pursue here, is that in the
ferropnictides, the single Fermi surface sheet we have studied
thus far is coupled to other sheets
 which control the $T_c$ suppression.  This appears plausible
 within the context of our microscopic spin-fluctuation pairing
 calculations\cite{ref:graseretal}, which show that for
 extended-$s$ type states, while the $\beta$ sheets appear to have
weak nodes, the $\alpha$ sheets are more isotropic for the
extended-$s$ wave solutions. %In addition, the pairing weight on
%the inner $\alpha$ sheet is substantially larger than on any of
%the other sheets (Fig. \ref{fig1}), as also found in ARPES
%experiments on Ba-122\cite{ref:Ding,ref:Evtushinsky}.
The pairing on the $\alpha$ and $\beta$ sheets is strongly coupled
by the $\pi,0$ (unfolded zone) scattering processes which dominate
the spin-fluctuation spectrum. To suppress the critical
temperature, it is clear that not only the $\beta$ condensate, but
also the $\alpha$ condensate must be suppressed. The $\alpha$
condensate may therefore be expected to act as a reservoir which
maintains the critical temperature (provided interband scattering
by disorder is relatively weak), while the $\beta$ condensate is
nodal and therefore dominates low-temperature and low-energy
properties.

\begin{figure}[]
\includegraphics[width=.85\columnwidth]{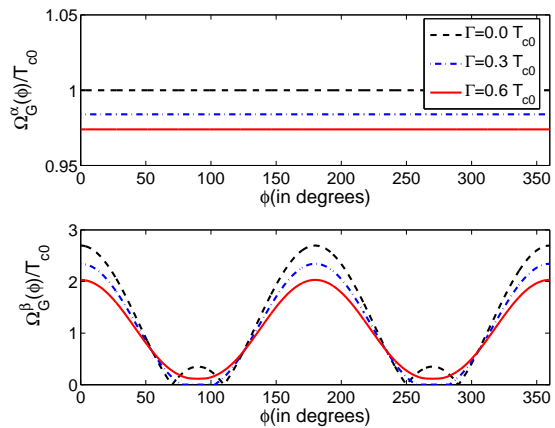}
\caption{(Color online)  Spectral gaps $\Omega_G(\phi)/T_{c0}$ vs.
$\phi$ for the $\alpha$ (top) and $\beta$ (bottom) sheets for the
2-band model. Here the pairing parameters
$V_{11}=1.0,V_{12}=-0.6,V_{22}=$1.5 were chosen so that
$r_{eff}=1.3$ is the same as in the 1-band case. The densities of
states ratio was taken as $N_2(0)/N_1(0)=1.25$. Scattering rates
for intraband scattering correspond to $\Gamma\equiv n_i \pi
N_1(0) |U_{11}|^2 =0$ (dashed), 0.3 (dashed-dotted), 0.6 (solid).
 }
\label{fig7}
\end{figure}

To test this hypothesis within the philosophy of the current
paper, we take the above model and add to it a single additional
sheet with constant pairing amplitude, and couple the two sheets
by constant  pairing potentials, leading to a total pairing
interaction.

\begin{eqnarray}
V(\k,\k') &=& V_1\Phi_1(\k)\Phi_1(\k') + V_2
\Phi_2(\k)\Phi_2(\k')\nonumber \\&&
+V_{12}[\phi_1(\k)\phi_2(\k')+\phi_2(\k)\phi_1(\k')],
\end{eqnarray}
where $\phi_1(\k)=1+r\cos 2\phi$ is understood as before to
describe states with $\k$ on the $\beta$ sheet, whereas
$\phi_2(\k)=1$ for $\k$ on the $\alpha$ sheet.  Note that $V_{12}$
is chosen of opposite sign to $V_1,V_2$ so as to induce a
sign-changing order parameter between the two sheets.

\begin{figure}[]
\includegraphics[width=.85\columnwidth]{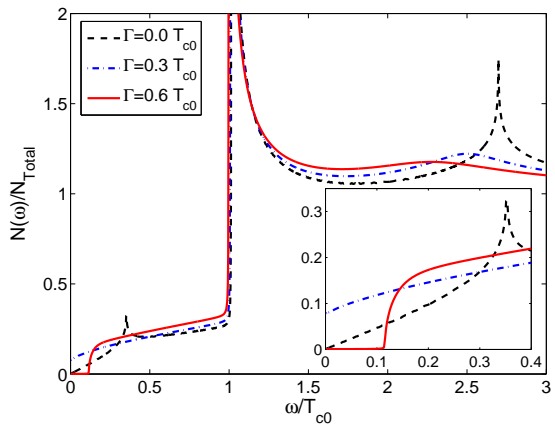}
\caption{(Color online)  Density of states $N(\omega)/N_{Total}$ vs.
$\omega/T_{c0}$ for 2-band system with $r_{eff}=1.3$.  Same
parameters and line types as in Fig. \ref{fig7}.Insert: expanded
low-energy region.}
\label{fig8}
\end{figure}

We now consider the scattering from nonmagnetic impurities.  As
discussed by Refs.
\onlinecite{ref:Mazindisorder,ref:Kontanidisorder,ref:YBang}, it
is convenient to parameterize the scattering potential in terms of
an amplitude  $U_{22}$ describing scattering $\k\rightarrow\k'$ on
the $\alpha$ sheet,  similarly $U_{11}$ on the $\beta$ sheet, and
$U_{12,21}$ between the two, such that the Born self-energy
becomes
\begin{eqnarray}
{\underline\Sigma}(\k\in 1,\omega) & = & {n_I} \, \left[
\sum_{\k'\in 1} |U_{11}|^2 {\underline G}(\k',\omega)\right.
\nonumber\\ && \left.  + \sum_{\k'\in 2} |U_{12}|^2 {\underline
G}(\k',\omega) \right],
\end{eqnarray}

\noindent and similarly for $\k\in 2$.  To compare the 1-band and
2-band cases, we now choose pairing parameters such that the order
parameter on the $\beta$ sheet is nearly identical to that we had
in the 1-band case above with $r_{eff}\equiv
(\Delta_{max}-\Delta_{min})/(\Delta_{max}+\Delta_{min})=1.3$. This
is illustrated as the pure spectral gap $\Omega_G$ in Fig.
\ref{fig7}, where the corresponding isotropic gap on the $\alpha$
sheet is also shown. As disorder is added, the nodes are removed
from the $\beta$ sheet, as before, and a full gap created. At the
same time the $\alpha$ gap is only very slightly suppressed in
this process. These features are evident also in the total DOS
shown in Fig. \ref{fig8} for the same scattering parameters.

\begin{figure}[]
\includegraphics[width=.85\columnwidth]{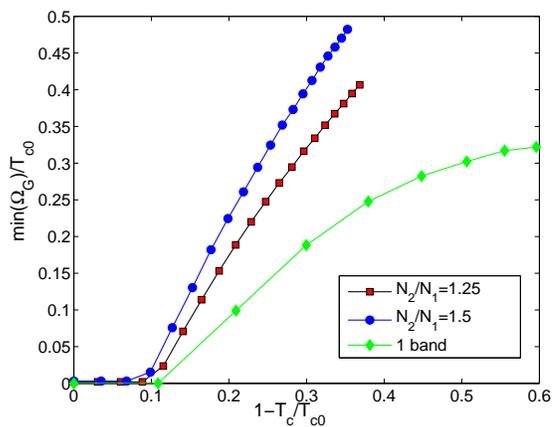}
\caption{(Color online)  Normalized spectral gap ${\rm
min}\Omega_G(\phi)/T_{c0}$ vs. $T_c$ suppression $1-T_c/T_{c0}$
for various impurity concentrations comparing 1-band and 2-band
models.  Parameters are chosen such that $r_{eff}=1.3$ on the $1$
($\beta$) sheet for the pure superconducting state in all cases.
Diamonds: 1-band model with $T_c/T_{c0}$ and $\Omega_G/T_{c0}$
taken from Figs. \ref{fig2} and \ref{fig3}, respectively. Squares:
two-band model with $V_{11}=1.0,V_{12}=-0.6,V_{22}=$1.5, with
densities of states ratio $N_2(0)/N_1(0)=1.25$ and 1 sheet
anisotropy parameter $r=1.76$. Circles: two-band model with same
$V_{11}=1.0,V_{12}=-0.6,V_{22}=$1.5, but with densities of states
ratio $N_2(0)/N_1(0)=1.5$ and $r=1.88$.
 }
\label{fig9}
\end{figure}

Our aim in this section is to see if within a two-band picture one
can understand qualitatively under what circumstances a
substantial spectral gap can be opened on the Fermi surface
without suppressing $T_c$ substantially in the simple 1-band
example. The naive hypothesis formulated above, that the presence
of a more isotropic gap on a second Fermi surface sheet is
sufficient to ``protect" $T_c$ against intraband scattering, is
not universally correct.  This is because the order parameters on
the two sheets are strongly coupled by the interband pair
interaction. To illustrate which aspects of the problem are
important for the relative proportion of gap creation relative to
$T_c$ suppression, we compare in Fig. \ref{fig9} the gap created
by a certain amount of disorder versus the corresponding
suppression of $T_c$, for the 1-band case and two  2-band cases
where the effective 1 sheet anisotropy $r_{eff}$ is held fixed,
but the ratio of the densities of states on the two sheets are
varied. It is seen that the critical temperature is rendered more
robust when the density of states $N_2(0)$, which controls the
pairing weight on  sheet 2 ($\alpha$), is increased.  Any effect
which enhances the nodeless  sheet 2 pairing weight makes $T_c$
less susceptible to intraband scattering.  Thus we conclude that
the amount of $T_c$ suppression associated with gap creation
depends on the details of the situation, and can become quite
small.

%On the other hand, when one examines the $T_c$ suppression, we see
%that scattering rates required to create gaps comparable to those
%in Figs. \ref{fig3} and \ref{fig4} suppress $T_c$ much more
%slowly.  Compare for example the $\Gamma=1.0T_{c0}$ curve in Fig.
%\ref{fig4}, where a spectral gap of about 0.22$T_{c0}$   is
%observed, corresponding to a 30\% suppression of $T_c$ seen in
%Fig. \ref{fig3}.   In the 2-band case, the same gap (Fig. \ref{8})
%corresponds to a $T_c$ suppression of only x\%.  In Fig.
%\ref{fig9}, we compare the $T_c$ suppression of the 1- and 2-band
%cases directly under various assumptions about the type of
%scattering.  In the above estimate, we have set interband
%scattering to zero, producing a slower suppression, as seen in the
%Figure.  On the other hand, a more isotropic scatterer will cause
%interband scattering, and  suppress $T_c$ dramatically due to the
%sign-changing character of the order
%parameter\ref{ref:Mazindisorder,ref:Kontanidisorder}.

The true Fermi surface structure is of course more complicated
than the 2-sheet model considered here, so it is interesting to
ask whether results from the more complete model show the desired
effect. We show in Fig. 10 calculations for the spectral gap in
the extended $s$-wave $(A_{1g})$ state obtained  from the
microscopic calculations of Ref. \onlinecite{ref:graseretal} for
various values of intraband disorder scattering. The node lifting
phenomenon on the $\beta$ sheets is clearly observed.  In Fig.
\ref{fig11}, we show that relatively little $T_c$ suppression
accompanies this realistic case; significant gaps of order
$0.1-0.2 T_c$ are obtained for $1-T_c/T_{c0}$ of only $\sim10\%$.

\begin{figure}[]
\includegraphics[width=1.1\columnwidth]{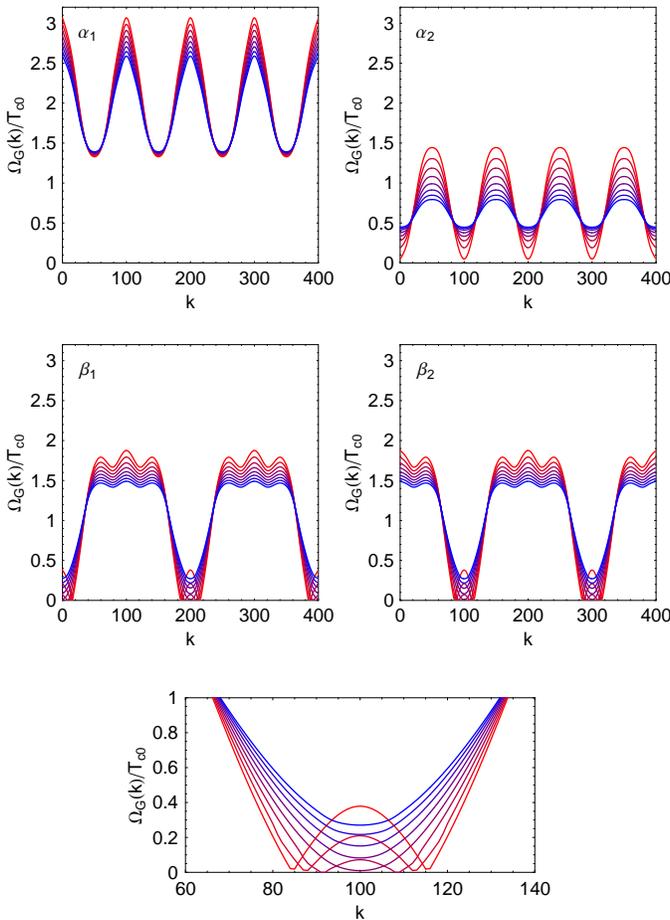}
%\\
%\includegraphics[width=.95\columnwidth]{SpectralGap_different_ni_detail_beta2.eps}
\caption{(Color online)  Spectral gap vs. Fermi surface arc length
in arbitrary units on each of four Fermi sheets ($\alpha_1$,
$\alpha_2$, $\beta_1$ and $\beta_2$) for realistic spin
fluctuation model of Fig. 17 of Ref. \onlinecite{ref:graseretal}.
Curves correspond to an arbitrary range of impurity
concentrations, from red (clean) to blue (dirty).  Bottom panel:
detail of nodal region of $\beta$ sheets.
 }
\label{fig10}
\end{figure}

\begin{figure}[]
\includegraphics[width=1.1\columnwidth]{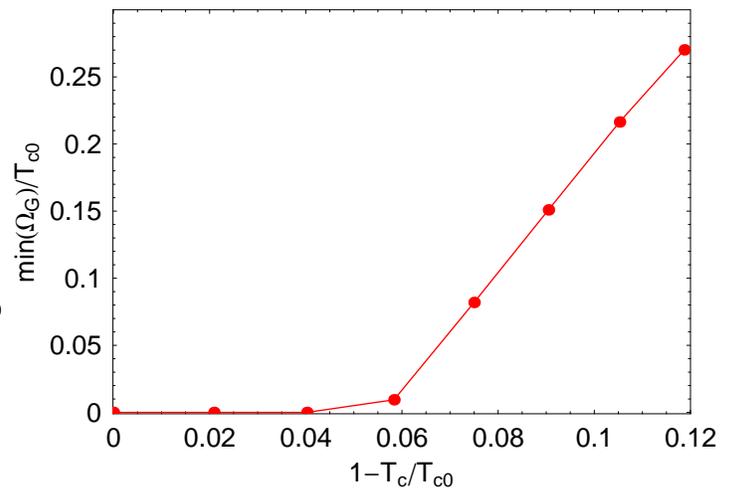}
\caption{(Color online) ${\rm min}\,\Omega_G/T_{c0}$ vs.
$1-T_c/T_{c0}$ for same disorder parameters as Fig. \ref{fig10}.
 }
\label{fig11}
\end{figure}

\section{Conclusions}
\label{sec:concl}

There is now  evidence that the superfluid density has a $T$
dependence consistent with a fully developed gap in some samples,
while power laws in $T$, including linear $T$, have been reported
in others.  It is possible that parameters related to the
electronic structure of the pure state  tune the various materials
such that different superconducting ground states are realized.
This is ``natural" due to the proximity of these systems to a
situation where nearly circular Fermi surface sheets nest
perfectly, in which case spin fluctuation theory predicts a
degeneracy between the extended-$s$ ($A_{1g}$)  and $d$ ($B_{1g}$)
states.

 Here we have explored an
alternative possibility which assumes that there are nodes in an
extended-$s$-wave ($A_{1g}$) state for the ideal ferropnictide
material, but that these nodes are lifted by small momentum
disorder scattering. In this regard, it is interesting that the
stoichiometric ferropnictide LaFePO (with a mean free path of more
than $10^2$nm and capable of supporting dHvA oscillations)
displays a linear $T$ dependence of the superfluid density at low
temperatures.  While many mechanisms can give rise to a power law
$T$ dependence in the superfluid density, we are aware of only one
explanation for a linear-$T$ law, namely the existence of nodes in
the pure state. On this basis we speculate that the exponential
behaviors observed in other materials are due to disorder.   On
the other hand, evidence for the opposite trend has been reported
in the Ba$_{1-x}$K$_x$Fe$_2$As$_2$ system, where the cleanest
crystals appear to exhibit exponential $T$ dependence, whereas
more disordered samples exhibit a low temperature dependence which
mimics a power law.

%However, this proximity is impossible to quantify, and it is
%useful to explore alternatives which involve a single primary
%pairing channel in all or most of the ferropnictides.  In this
%regard we consider it significant that the cleanest ferropnictide
%samples prepared to date, LaFePO, displays a superfluid density
%with linear $T$ dependence.

We have therefore explored a scenario in which weak nodes in a
sign-changing $s_{+/-}$ state on the $\beta$ sheets of the
ferropnictides are ``lifted" by nonmagnetic disorder.  Qualitative
aspects of this phenomenon are obtainable already within a simple
1-band model where weak pointlike scatterers are accounted for.
With increasing disorder, the nodes disappear and are replaced by
a fully developed gap and an activated temperature dependence of
$\rho_s$.  Thus one could imagine that the entire class of
ferropnictide materials has intrinsic nodes in the ground state
order parameter of the analog pure system, which then disappear in
most of the doped, much dirtier materials.  We have shown that in
the simplest model the lifting of the nodes corresponds to a
significantly larger pairbreaking rate and concomitant $T_c$
suppression than observed experimentally, but that this
undesirable aspect of the theory  is substantially eliminated by
the addition of additional bands with more isotropic pair state,
as found in the microscopic theory\cite{ref:graseretal}.

The justification for our neglect of interband scattering in most
of this work is based on the empirical robustness of $T_c$ to
differences in sample quality, suggesting that interband
scattering in the sign-changing $s$ state is negligible.  In
addition, Yukawa-type models of realistic impurity potentials with
finite range have small large-$q$ amplitudes.  Nevertheless, for a
quantitative description of these materials both types of
scattering should be considered, and a more microscopic
understanding of the types of potentials introduced, e.g. by a K
replacing Ba,Sr out of the FeAs plane and by a Co substituting
directly for an Fe would be very useful.

The simple theory presented in Section \ref{sec:model} is actually
applicable to a state with arbitrary anisotropy parameter $r$,
allowing us to describe the pure $d$-wave state for $r\gg 1$, as
well as a state with deep minima of the order parameter
(``quasi-nodes"), but no true nodes for $r\lesssim 1$ as found,
e.g. in Ref. \onlinecite{ref:Wang}.  Thus we emphasize that it is
possible that small differences in the electronic structure of
different materials may give rise to slightly different gap
structures, and even different symmetry
classes\cite{ref:graseretal} in different materials.  The
implication of current ARPES experiments, that the order parameter
anisotropy around the various Fermi surface sheets is absent or
very weak is, however, difficult to reconcile with experimental
data which indicate low energy excitations, and a disorder-based
explanation of the kind given here, where nodes or quasi-nodes are
lifted by momentum averaging, seems to us the most likely way to
understand the existing data.  This picture will have implications
for other properties, including transport properties, as discussed
in Ref. \onlinecite{ref:putikka} for the 1-band situation, but
there may be interesting modifications peculiar to the multiband
case. Work on this direction is in progress.

We note in closing that an alternative approach takes the point of
view that the intrinsic material is fully gapped with a
sign-changing $s$-wave state, and that the inter-Fermi surface
scattering leads to the appearance of low-temperature power law
behavior\cite{ref:MazinNMR,ref:ChubukovNMR,ref:Vorontsov}. Clearly
systematic electron irradiation or some other source of disorder
which does not change the doping of the system would be a very
important experimental way to distinguish this proposal from our
own.

 \acknowledgements  The authors are grateful for  useful communications
 with D.A. Bonn, J. Bobowski, and A. Carrington.  Research was
 partially supported by DOE DE-FG02-05ER46236 (PJH), and the
Deutscheforschungsgemeinschaft (SG). TAM, DJS, and PJH acknowledge
the Center for Nanophase Materials Science, which is sponsored at
Oak Ridge National Laboratory by the Division of Scientific User
Facilities, U.S. Department of Energy.


\begin{thebibliography}{99}

\bibitem{ref:tsuei} C.C.~Tsuei and J.R.~Kirtley,  Rev. Mod. Phys. {\bf 72}, 969 (2000).

\bibitem{ref:kamihara} Y. Kamihara, T. Watanabe, M. Hirano, and  H. Hosono,
 J. Am. Chem. Soc. {\bf 130}, 3296 (2008).

 \bibitem{ref:Hashimoto} K. Hashimoto, T. Shibauchi, T. Kato, K. Ikada,
 R. Okazaki, H. Shishido, M. Ishikado, H. Kito, A. Iyo, H. Eisaki, S. Shamoto, and Y. Matsuda arXiv:0806.3149.


 \bibitem{ref:Malone} L. Malone, J.D. Fletcher, A. Serafin, A. Carrington, N.D. Zhigadlo,
 Z. Bukowski, S. Katrych, and J. Karpinski,  arXiv 0807.0876.

 \bibitem{ref:Martin}C. Martin, R. T. Gordon, M. A. Tanatar, M. D. Vannette,
  M. E. Tillman, E. D. Mun, P. C. Canfield, V. G. Kogan, G. D. Samolyuk, J. Schmalian, and R. Prozorov,  ArXiv:0807.0876

 \bibitem{ref:Hashimoto2}K. Hashimoto etal. arXiv:0810.3506.

 \bibitem{ref:Gordon} R. T. Gordon, N. Ni, C. Martin, M. A. Tanatar, M. D. Vannette, H. Kim, G. Samolyuk, J. Schmalian,
  S. Nandi, A. Kreyssig, A. I. Goldman, J. Q. Yan, S. L. Bud'ko, P. C. Canfield, R.
  Prozorov, arXiv:0810.2295.

  \bibitem{ref:Gordon2} R. T. Gordon, C. Martin, H. Kim, N. Ni, M. A. Tanatar, J. Schmalian,
  I. I. Mazin, S. L. Bud'ko, P. C. Canfield, R. Prozorov,
  arXiv:0812.3683.


 \bibitem{ref:Fletcher} J.D. Fletcher, A. Serafin, L. Malone, J. Analytis, J-H Chu, A.S. Erickson, I.R. Fisher, A.
 Carrington, arXiv:0812.3858.


\bibitem{ref:Zhao}L. Zhao et al Chin. Phys. Lett. {\bf
             25}, 4402 (2008).

 \bibitem{ref:Ding} H.~Ding, P.~Richard, K.~Nakayama, T.~Sugawara, T.~Arakane, Y.~Sekiba, A.~Takayama,
             S.~Souma, T.~Sato, T.~Takahashi, Z.~Wang, X.~Dai, Z.~Fang, G.F.~Chen, J.L.~Luo, N.L.~Wang,
             Europhys. Lett. {\bf 83}, 47001 (2008).

 \bibitem{ref:Kondo} T.~Kondo, A.F.~Santander-Syro, O.~Copie, C.~Liu, M.E.~Tillman, E.D.~Mun,
             J.~Schmalian, S.L.~Bud'ko, M.A.~Tanatar, P.C.~Canfield, A.~Kaminski,
             Phys. Rev. Lett. {\bf 101}, 147003 (2008).

 \bibitem{ref:Evtushinsky} D.V.~Evtushinsky, D.S.~Inosov, V.B.~Zabolotnyy, A.~Koitzsch, M.~Knupfer,
             B.~Buchner, G.L.~Sun, V.~Hinkov, A.V.~Boris, C.T.~Lin, B.~Keimer, A.~Varykhalov, A.A.~Kordyuk, S.V.~Borisenko,
             arXiv:0809.4455.


\bibitem{ref:Nakayama}K. Nakayama, T. Sato, P. Richard, Y.-M. Xu, Y. Sekiba, S. Souma,
G. F. Chen, J. L. Luo, N. L. Wang, H. Ding, T. Takahashi,
arXiv:0812.0663.

\bibitem{ref:Hasan} L. Wray, D. Qian, D. Hsieh, Y. Xia, L. Li, J.G. Checkelsky, A. Pasupathy, K.K. Gomes,
C.V. Parker, A.V. Fedorov, G.F. Chen, J.L. Luo, A. Yazdani, N.P.
Ong, N.L. Wang, M.Z. Hasan, arXiv: 0812.2061.

\bibitem{ref:RKlingeler} R. Klingeler, N.~Leps, I.~Hellmann, A.~Popa, C.~Hess,
             A.~Kondrat, J.~Hamann-Borrero, G.~Behr, V.~Kataev, and B.~Buechner,
             arXiv:0808.0708.

\bibitem{ref:Grafe} H.-J.~Grafe, D.~Paar, G.~Lang, N.J.~Curro,
             G.~Behr, J.~Werner, J.~Hamann-Borrero, C.~Hess, N.~Leps,
             R.~Klingeler, and B.~Buchner, Phys. Rev. Lett. {\bf 101}, 047003 (2008).

\bibitem{ref:Ahilan} K.~Ahilan, F.L.~Ning, T.~Imai, A.S.~Sefat, R.~Jin, M.A.~McGuire,
B.C.~Sales, D.~Mandrus, Phys. Rev. B {\bf 78}, 100501(R) (2008).

\bibitem{ref:Nakai} T.Y. Nakai et al., J. Phys. Soc. Jpn. {\bf
77}, 073701 (2008).

\bibitem{ref:Shan} L.~Shan, Y.~Wang, X.~Zhu, G.~Mu, L.~Fang, C.~Ren and H.-H.~Wen,
             Europhys. Lett. {\bf 83}, 57004 (2008).

\bibitem{ref:Chien} T.Y. Chien, Z. Tesanovic, R.H. Liu, X.H. Chen,
and C.L. Chien, Nature {\bf 453}, 1224 (2008).

\bibitem{ref:Daghero} D. Daghero et al., arXiv:0812.1141.


\bibitem{ref:Gonnelli} R.S. Gonnelli et al., arXiv:0807.3149.


 \bibitem{ref:Kuroki} K.~Kuroki, S.~Onari, R.~Arita, H.~Usui, Y.~Tanaka, H.~Kontani
             and H.~Aoki, Phys. Rev. Lett. {\bf 101}, 087004 (2008).

\bibitem{ref:Wang} F.~Wang, H.~Zhai, Y.~Ran, A.~Vishwanath and D.-H.~Lee,
             arXiv:0807.

\bibitem{ref:Mazin_exts}I. I. Mazin, D. J. Singh, M. D. Johannes, and M. H. Du, Phys. Rev. Lett. 101, 057003
(2008).

\bibitem{ref:Chubukov_exts} A.V. Chubukov, D. Efremov, and I.
Eremin, Phys. Rev. {\bf B 78}, 134512 (2008).

\bibitem{ref:graseretal} S. Graser, T. A. Maier, P. J. Hirschfeld, and D. J.
Scalapino,  arXiv:0812.0343.

\bibitem{ref:Cao} C.~Cao, P.J.~Hirschfeld, H.-P.~Cheng, Phys. Rev. B {\bf 77}, 220506(R) (2008).

\bibitem{ref:Borkowski} L. S. Borkowski and P.J. Hirschfeld, Phys. Rev.
B49, 15404 (1994).

\bibitem{ref:Fehrenbacher}  R. Fehrenbacher and M. R. Norman,
 Phys. Rev. B 50, 3495 (1994).

\bibitem{ref:Muzikar} G. Preosti and P. Muzikar, Phys. Rev. B 54, 3489
(1996).



\bibitem{ref:Mazindisorder}  A.A. Golubov and I.I. Mazin, Phys. Rev. B 55, 15146 (1997).

\bibitem{ref:Kontanidisorder} Y. Senga and H. Kontani,
arXiv:0809.0374;     arXiv:0812.2100.  One of these is J. Phys.
Soc. Jpn. {\bf 77}, 113710 (2008).

\bibitem{ref:YBang} Y. Bang, H.-Y. Choi, and H. Won, arXiv:
0808.3473.

\bibitem{ref:Vorontsov} A.B. Vorontsov,  M.G. Vavilov, and A.V.
Chubukov, arXiv 0901.0719.

\bibitem{ref:MazinNMR} D. Parker, O.V. Dolgov, M.M. Korshunov, A.A. Golubov, and I.I.
Mazin, arXiv:0807.3729.

\bibitem{ref:ChubukovNMR} A.V. Chubukov, D. Efremov, and I.
Eremin, arXiv:0807.3735.


\bibitem{ref:Chubukov_rhos} A.B. Vorontsov, M.G. Vavilov, A.V.
Chubukov, arXiv:arXiv:0901.0719.

\bibitem{ref:Kadanoff} D. Markowitz and L. P. Kadanoff, Phys. Rev. 131,
563 (1963).

\bibitem{ref:AG} A.A. Abrikosov and L.P. Gor'kov,
 Zh. Eksp. Teor.  Fiz  {\bf 39}, 1781(60) [Sov. Phys. JETP
{\bf 12}, 1243(1961)].

\bibitem{ref:Skalski}S. Skalski, O. Betbeder-Matibet, and P. R. Weiss,
Phys. Rev. {\bf 136}, A1500 (1964).

\bibitem{ref:HWE88} P.J. Hirschfeld, P. W\"olfle and D.
Einzel, Phys. Rev.  B 37 83, (1988).

\bibitem{ref:footnote1} We have averaged the $xx$ and $yy$
components to simulate the ferropnictide with order parameters on
$\beta_1$ and $\beta_2$ sheets rotated by $\pi/2$ in local
coordinates with respect to one another.

\bibitem{ref:footnote2}  The approximate factor of 2 difference
between the kink temperature and the antinodal order parameter
energy energy arises from the width of the thermal quasiparticle
distribution.

\bibitem{ref:Gross} F. Gross, B.S. Chandrasekhar, D. Einzel, P.J. Hirschfeld, K.
Andres, H. R. Ott, J. Beuers, Z. Fisk and J.L. Smith, Z. Physik  B
64, 175 (1986).

\bibitem{ref:Koztin} I. Koztin and A.J. Leggett, Phys. Rev. Lett.
{\bf 79}, 135 (1997).

\bibitem{ref:KamiharaP} Y. Kamihara, et al. J. Amer. Chem. Soc.
{\bf 128}, 10012 (2006).


\bibitem{ref:Carrington1} A.I. Coldea et al., Phys. Rev. Lett.
{\bf 101}, 216402 (2008).

\bibitem{ref:felds} P.J. Hirschfeld
and N. Goldenfeld,  Phys. Rev. B. 48, 4219(1993).





\bibitem{ref:putikka} L. S. Borkowski,  P.J. Hirschfeld, and
W.O. Putikka, Phys. Rev. B52, 3856 (1995).

%\bibitem{ref:Malaeb}
%
%W. Malaeb, T. Yoshida, T. Kataoka, A. Fujimori, M. Kubota, K. Ono,
%H. Usui, K. Kuroki, R. Arita, H. Aoki, Y. Kamihara, M. Hirano and
%H. Hosono J. Phys. Soc. Jpn. 77 (2008) 093714.
%
%\bibitem{ref:Lebegue} S.~Leb\`egue, Phys. Rev. B {\bf 75}, 035110 (2007).
%
%\bibitem{ref:Singh} D.J.~Singh and M.-H.~Du,  Phys. Rev. Lett. {\bf 100}, 237003 (2008).
%
%\bibitem{ref:Cao} C.~Cao, P.J.~Hirschfeld, H.-P.~Cheng, Phys. Rev. B {\bf 77}, 220506(R) (2008).
%
%\bibitem{ref:Andersenpc} O.K.~Andersen, private communication.
%
%\bibitem{ref:Dong} J.~Dong, H.J.~Zhang, G.~Xu, Z.~Li, G.~Li, W.Z.~Hu, D.~Wu,
%             G.F.~Chen, X.~Dai, J.L.~Luo, Z.~Fang and N.L.~Wang,
%             Europhys. Lett. {\bf 83}, 27006 (2008).
%
%\bibitem{ref:delaCruz} C.~de~la~Cruz, Q.~Huang, J.W.~Lynn, J.~Li,
%             W.~Ratcliff, J.L.~Zarestky, H.A.~Mook, G.F.~Chen, J.L.~Luo,
%             N.L.~Wang, and P.~Dai, {\it Nature}  {\bf 453}, 899 (2008).
%
%\bibitem{ref:Zhao} J.~Zhao, Q.~Huang, C.~de~la~Cruz, S.~Li, J.W.~Lynn, Y.~Chen,
%             M.A.~Green, G.F.~Chen, G.~Li, Z.~Li, J.L.~Luo, N.L.~Wang, and P.~Dai,
%             Nature Materials {\bf 7}, 953 (2008).
%
%\bibitem{ref:Luetkens}  H.~Luetkens, H.-H.~Klauss, M.~Kraken, F.J.~Litterst, T.~Dellmann, R.~Klingeler,
%             C.~Hess, R.~Khasanov, A.~Amato, C.~Baines, J.~Hamann-Borrero, N.~Leps, A.~Kondrat, G.~Behr,
%             J.~Werner, B.~Buechner, arXiv:0806.3533.
%
%\bibitem{ref:Drew} A.J.~Drew, Ch.~Niedermayer, P.J.~Baker, F.L.~Pratt, S.J.~Blundell, T.~Lancaster,
%             R.H.~Liu, G.~Wu, X.H.~Chen, I.~Watanabe, V.K.~Malik, A.~Dubroka,
%             M.~R\"ossle, K.W.~Kim, C.~Baines and C.~Bernhard, arXiv:0807.4876.
%

%
%\bibitem{ref:Qi} X.-L.~Qi, S.~Raghu, C.-X.~Liu, D.J.~Scalapino and S.-C.~Zhang,
%             arXiv:0804.4332.
%
%\bibitem{ref:Barzykin} V.~Barzykin and L.P.~Gorkov, arXiv:0806.1933.
%
%\bibitem{ref:Bang} Y.~Bang and H.-Y.~Choi,
%             Phys. Rev. B {\bf 78}, 134523 (2008).
%
%\bibitem{ref:Yao} Z.-J.~Yao, J.-X.~Li and Z.D.~Wang, arXiv:0804.4166.
%
%\bibitem{ref:Sknepnek} R.~Sknepnek, G.~Samolyuk, Y.~Lee, B.N.~Harmon and J.~Schmalian,
%            arXiv:0807.4566.
%


%\bibitem{ref:Chubukov} A.V.~Chubukov, D.~Efremov and I.~Eremin,
%             Phys. Rev. B {\bf 78}, 134512 (2008).
%
%\bibitem{ref:Mu} G.~Mu, X.-Y.~Zhu, L.~Fang, L.~Shan, C.~Ren and H.-H.~Wen,
%             Chin. Phys. Lett. {\bf 25}, 2221 (2008); arXiv:0803.0928.
%
%\bibitem{ref:Matano} K.~Matano, Z.A.~Ren, X.L.~Dong, L.L.~Sun, Z.X.~Zhao, G.~Zheng,
%             Europhys. Lett. {\bf 83}, 57001 (2008).
%
%\bibitem{ref:Mukuda} H.~Mukuda, N.~Terasaki, H.~Kinouchi, M.~Yashima, Y.~Kitaoka, S.~Suzuki,
%             S.~Miyasaka, S.~Tajima, K.~Miyazawa, P.M.~Shirage, H.~Kito, H.~Eisaki, A.~Iyo, arXiv:0806.3238.
%
%\bibitem{ref:Nakai} Y.~Nakai, K.~Ishida, Y.~Kamihara, M.~Hirano, H.~Hosono, arXiv:0804.4765.
%
%\bibitem{ref:ChenYY} T.Y.~Chen, Z.~Tesanovic, R.H.~Liu, X.H.~Chen, C.L.~Chien,
%             {\it Nature} {\bf 453}, 1224 (2008);
%             K.A.~Yates, L.F.~Cohen, Z.-A.~Ren, J.~Yang, W.~Lu, X.-L.~Dong, Z.-X.~Zhao,
%             Supercond. Sci. Technol. {\bf 21}, 092003 (2008).
%
%\bibitem{ref:Kondo} T.~Kondo, A.F.~Santander-Syro, O.~Copie, C.~Liu, M.E.~Tillman, E.D.~Mun,
%             J.~Schmalian, S.L.~Bud'ko, M.A.~Tanatar, P.C.~Canfield, A.~Kaminski,
%             Phys. Rev. Lett. {\bf 101}, 147003 (2008).
%
%%\bibitem{ref:Jia} X.~Jia, H.~Liu, W.~Zhang, L.~Zhao, J.~Meng, G.~Liu, X.~Dong, G.F.~Chen,
%%            J.L.~Luo, N.L.~Wang, Z.A.~Ren, W.~Yi, J.~Yang, W.~Lu, G.C.~Che, G.~Wu, R.H.~Liu,
%%             X.H.~Chen, G.~Wang, Y.~Zhou, Y.~Zhu, X.~Wang, Z.~Zhao, Z.~Xu, C.~Chen, X.J.~Zhou,
%%             Chin. Phys. Lett. {\bf 25}, 3765 (2008); arXiv:0806.0291.
%%
%%\bibitem{ref:ChenH} H.~Chen, Y.~Ren, Y.~Qiu, W.~Bao, R.H.~Liu, G.~Wu, T.~Wu, Y.L.~Xie,
%%             X.F.~Wang, Q.~Huang, X.H.~Chen, arXiv:0807.3950.
%
%\bibitem{ref:Li} G.~Li, W.Z.~Hu, J.~Dong, Z.~Li, P.~Zheng, G.F.~Chen, J.L.~Luo, N.L.~Wang,
%             Phys. Rev. Lett. {\bf 101}, 107004 (2008).
%
%%\bibitem{ref:Christianson} A.D.~Christianson, E.A.~Goremychkin, R.~Osborn, S.~Rosenkranz, M.D.~Lumsden,
%%               C.D.~Malliakas, l.S.~Todorov, H.~Claus, D.Y.~Chung, M.G.~Kanatzidis, R.I.~Bewley, T.~Guidi,
%%               arXiv:0807.3932.
%

%
%
%\bibitem{ref:Martin} C.~Martin, R.T.~Gordon, M.A.~Tanatar, M.D.~Vannette, M.E.~Tillman, E.D.~Mun,
%             P.C.~Canfield, V.G.~Kogan, G.D.~Samolyuk, J.~Schmalian, R.~Prozorov, arXiv:0807.0876.
%
%\bibitem{ref:Hashimoto} K.~Hashimoto, T.~Shibauchi, S.~Kasahara, K.~Ikada, T.~Kato, R.~Okazaki,
%             C.J.~van~der~Beek, M.~Konczykowski, H.~Takeya, K.~Hirata, T.~Terashima, Y.~Matsuda,
%             arXiv:0810.3506.
%

%
%\bibitem{ref:LeeWen} P.A.~Lee and X.-G.~Wen, arXiv:0804.1739.
%
%\bibitem{ref:anisimov} V.I.~Anisimov, Dm.M.~Korotin, M.A.~Korotin,
%             A.V.~Kozhevnikov, J.~Kunes, A.O.~Shorikov, S.L.~Skornyakov,
%             and S.V.~Streltsov, arXiv:0810.2629.
%
%\bibitem{ref:oles} A.M.~Ol\'es, Phys. Rev. B {\bf 28}, 327 (1983).
%
%\bibitem{ref:Takimoto} T.~Takimoto, T.~Hotta, and K.~Ueda,
%             Phys. Rev. B {\bf 69}, 104504 (2004).
%
%\bibitem{ref:kubo} K.~Kubo, Phys. Rev. B {\bf 75}, 224509 (2007).
%
%%\bibitem{ref:GWu} G.~Wu, H.~Chen, T.~Wu, Y. L.~Xie, Y. J.~Yan, R. H.~Liu, X.
%%F.Wang, J. J.~Ying, and X. H.~Chen, arXiv:0806.4279.
%
%%\bibitem{ref:XFWang} X. F. Wang, T. Wu, G. Wu, H. Chen, Y. L. Xie, J. J. Ying, Y.
%%J. Yan, R. H. Liu, and X. H. Chen, arXiv:0806.2452.
%
%%\bibitem{ref:YQYan} Y.Q. Yan, A. Kreyssig, S. Nandi, N. Ni, S. L. Bud'ko, A.
%%Kracher, R. J. McQueeney, R. W. McCallum, T. A. Lograsso, A. I.
%%Goldman, and P. C. Canfield, arXiv:0806.2711.
%

%
%\bibitem{ref:DHLeelinearchi} G.-M.~Zhang, Y.-H.~Su, Z.-Y.~Lu, Z.-Y.~Weng, D.-H.~Lee,
%             T.~Xiang,  arXiv:0809.3874.
%
%\bibitem{ref:Mazin} I.I.~Mazin, D.J.~Singh, M.D.~Johannes and M.H.~Du,
%             Phys. Rev. Lett. {\bf 101}, 057003 (2008).
%
%\bibitem{ref:Bickers} N.E.~Bickers, D.J.~Scalapino and S.R.~White,
%             Phys. Rev. Lett. {\bf 62}, 961 (1989).
%
%\bibitem{ref:bardasis} A.~Bardasis, and J.R.~Schrieffer, Phys. Rev. {\bf 121}, 1050 (1961).
%
%\bibitem{CommentKuroki} Note: Even when taking the band parameters given in
%            Kuroki {\it et al.} and identical interaction parameters, we find a different Fermi surface
%            and pairing eigenvalues. Also the extended $s$ wave pairing function we find for these
%            parameters differes from the one found by Kuroki {\it et al}. In contrast to this work
%            Kuroki {\it et al.} use a dyamical susceptibility, but it seems unlikely that this could account for these differences.
%
%\bibitem{ref:Ning} F.~Ning, K.~Ahilan, T.~Imai, A.S.~Sefat, R.~Jin, M.A.~McGuire, B.C.~Sales,
%             and D.~Mandrus, J. Phys. Soc. Jpn. {\bf 77} 103705 (2008).
%
%\bibitem{ref:sczhang_spid} W.-C. Lee, S.-C. Zhang
%and C. Wu, arXiv 0810.5114.
%
%\bibitem{ref:joynt} K.A.~Musaelian, J.~Betouras, A.V.~Chubukov, and R.~Joynt,
%             Phys. Rev. B {\bf 53}, 3598 (1996).






\end{thebibliography}
\end{document}